\documentclass[aps,prb,twocolumn,superscriptaddress]{revtex4}
\usepackage{graphicx}
\usepackage{amssymb}

\begin{document}
\title{Electronic correlations and Hund's 
coupling effects in SrMoO$_3$ 
revealed by photoemission spectroscopy}

\author{H.~Wadati}
\email{wadati@ap.t.u-tokyo.ac.jp}
\homepage{http://www.geocities.jp/qxbqd097/index2.htm}
\affiliation{Department of Applied Physics and Quantum-Phase Electronics 
Center (QPEC), University of Tokyo, Hongo, Tokyo 113-8656, Japan} 

\author{K.~Yoshimatsu}
\affiliation{Department of Applied Chemistry, University of Tokyo,
Bunkyo-ku, Tokyo 113-8656, Japan}

\author{H.~Kumigashira}
\affiliation{Department of Applied Chemistry, University of Tokyo,
Bunkyo-ku, Tokyo 113-8656, Japan}

\author{M.~Oshima}
\affiliation{Department of Applied Chemistry, University of Tokyo,
Bunkyo-ku, Tokyo 113-8656, Japan}

\author{T.~Sugiyama}
\affiliation{Japan Synchrotron Radiation Research Institute, 
SPring-8, Hyogo 679-5198, Japan}

\author{E.~Ikenaga}
\affiliation{Japan Synchrotron Radiation Research Institute, 
SPring-8, Hyogo 679-5198, Japan}
 
 \author{A.~Fujimori}
\affiliation{Department of Physics, University of Tokyo, Bunkyo-ku, Tokyo 113-0033, Japan}

\author{J.~Mravlje}
\affiliation{Jozef Stefan Institute, Jamova 39, Ljubljana, Slovenia} 
\affiliation{Coll\`ege de France, 11 place Marcelin Berthelot, 75005 Paris, France}
\affiliation{Centre de Physique Th\'eorique, \'Ecole Polytechnique, CNRS,
91128 Palaiseau, France} 

\author{A.~Georges}
\affiliation{Coll\`ege de France, 11 place Marcelin Berthelot, 75005 Paris, France}
\affiliation{Centre de Physique Th\'eorique, \'Ecole Polytechnique, CNRS, 
91128 Palaiseau, France}
\affiliation{DPMC,
Universite de Gen\`eve, 24 quai Ernest-Ansermet, 1211 Gen\`eve 4,
Switzerland}
 
\author{A. Radetinac} 
\affiliation{Institut f\"{u}r Materialwissenschaft, Technische Universit\"{a}t 
Darmstadt, Petersenstrabe 23, 64287 Darmstadt, Germany}

\author{K. S. Takahashi}
\affiliation{RIKEN Center for Emergent Matter Science (CEMS), 
Wako 351-0198, Japan}

\author{M. Kawasaki}
\affiliation{Department of Applied Physics and Quantum-Phase Electronics 
Center (QPEC), University of Tokyo, Hongo, Tokyo 113-8656, Japan} 
\affiliation{RIKEN Center for Emergent Matter Science (CEMS), 
Wako 351-0198, Japan}

\author{Y. Tokura}
\affiliation{Department of Applied Physics and Quantum-Phase Electronics 
Center (QPEC), University of Tokyo, Hongo, Tokyo 113-8656, Japan} 
\affiliation{RIKEN Center for Emergent Matter Science (CEMS), 
Wako 351-0198, Japan}

\pacs{71.30.+h, 71.28.+d, 73.61.-r, 79.60.Dp}
 
\date{\today}
\begin{abstract}
We investigate the electronic structure of a perovskite-type Pauli 
paramagnet SrMoO$_3$ (${t_{2g}}^2$) thin film using 
hard x-ray photoemission spectroscopy and compare the results 
to the realistic calculations 
that combine the density functional theory within the local-density 
approximation (LDA) with the dynamical-mean field 
theory (DMFT).  Despite the clear signature of electron correlations 
in the electronic specific heat, the narrowing of the quasiparticle 
bands is not observed in the photoemission spectrum.  This is 
explained in terms of the characteristic effect of Hund's rule 
coupling for partially-filled $t_{2g}$ bands, which induces strong 
quasiparticle renormalization already for values of Hubbard 
interaction which are smaller than the bandwidth.  The interpretation 
is supported by additional model DMFT calculations including 
Hund's rule coupling, that show renormalization of low-energy 
quasiparticles without affecting the overall bandwidth. The 
photoemission spectra show additional spectral weight around $-2$ eV 
that is not present in the LDA+DMFT. We interpret this weight as a 
plasmon satellite, which is supported by measured Mo, Sr and 
Oxygen core-hole spectra that all show satellites at this energy. 
\end{abstract}
\pacs{71.30.+h, 71.28.+d, 79.60.Dp, 73.61.-r}
\maketitle
\section{Introduction}
Electron correlation in transition-metal oxides (TMOs) has been the
subject of extensive studies in recent decades~\cite{rev}.
Photoemission spectroscopy has made major contributions to a better 
understanding of electron correlation effects in those materials.
Perovskite-type SrVO$_3$ has been extensively studied as a
prototypical correlated system with the ${t_{2g}}^1$ configuration and
no magnetism~\cite{D1,IHInouePRL}.  The bulk spectrum of the V $3d$
bands was obtained by using soft x-ray (SX) photoemission spectroscopy
and it was found that the width of the coherent part (quasiparticle
band) $W^{\ast}$ is reduced to about half of that found in the
band-structure calculation $W_b$~\cite{Sekiyama}, that is,
  $W_b/W^{\ast}\sim 2$.  This is consistent with specific heat
measurements, which suggest $m^{\ast}/m_b={\gamma}/{\gamma}_b\sim 2$,
where $\gamma$ is the experimental specific heat coefficient,
$\gamma_b$ is the theoretical specific heat coefficient obtained from
the band-structure calculations, $m^{\ast}$ is the effective mass of
the quasiparticle, and $m_b$ is the bare band mass~\cite{IHInoue}. 
The photoemission spectra also displayed a clear lower Hubbard band. 
Such a photoemission signal with well-separated Hubbard bands 
and a narrow quasiparticle peak has become an icon of correlated electron materials.

Surprisingly, the situation in TMOs with more than one $d$ electron
has been found to be quite different.  Among such systems, SrRuO$_3$
(${t_{2g}}^4$) has attracted particular interest due to the metalicity
and ferromagnetism with $T_C \sim$ 160 K~\cite{SROtc}. 
Takizawa~{\it et al.} \cite{TakizawaSCRO1} 
obtained a bulk Ru $4d$ spectrum 
of SrRuO$_3$ through SX photoemission studies 
of {\it in situ} prepared thin films, and found 
that the bandwidths found in the experimental bulk spectrum agree with 
those found in the band-structure calculation, that is, 
$W_b/W^{\ast}\sim 1$. This result, however, now
 does not match $m^{\ast}/m_b={\gamma}/{\gamma}_b \sim 4$ found from
specific heat measurements \cite{PBallen,Jokamoto}, which lead
Takizawa~{\it et al.}  to a conclusion that the genuine coherent part
with the renormalized electron mass exists only in the vicinity of the
Fermi level ($E_F$).  In another ${t_{2g}}^4$ system Sr$_2$RuO$_4$,
which is a layered superconductor \cite{maenonat}, the situation is
quite similar to SrRuO$_3$ in the sense that $W_b/W^{\ast}\sim 1$ but
$m^{\ast}/m_b={\gamma}/{\gamma}_b \sim 4$ 
\cite{maenonat, TOguchi, SekiyamaSr2RuO4}. 
These facts are summarized in Table \ref{tab1}. 
The absence of pronounced Hubbard bands in the photoemission spectra of ruthenates 
led some of the researchers in the field to adopt the extreme view that electronic
correlations are altogether absent or negligible in SrRuO$_3$ and CaRuO$_3$ compounds~\cite{JMmaiti}.

\begin{table}
\begin{center}
\caption{Effects of electron correlation 
in transition-metal oxides. The values of 
$\gamma$, $\gamma/\gamma_b$ $(=m^{\ast}/m_b)$, 
  and $W_b/W^{\ast}$ are given. The values 
without reference numbers are from this work.}
\begin{tabular}{cccc}
\hline
\hline
 & $\gamma$ & $\gamma/\gamma_b$ & $W_b/W^{\ast}$ \\
 & (mJ/K$^{2}$ mol) & $(=m^{\ast}/m_b)$&  \\
\hline
SrVO$_3$ (${t_{2g}}^1$) & 8.182 \footnotemark[1] 
& $\sim$ 2 \footnotemark[1] 
& $\sim$ 2 \footnotemark[2] \\ 
SrRuO$_3$ (${t_{2g}}^4$) & 36.3 \footnotemark[3] 
& $\sim$ 4  \footnotemark[3] 
& $\sim$ 1 \footnotemark[4] \\ 
Sr$_2$RuO$_4$ (${t_{2g}}^4$) & 39 \footnotemark[5] 
& $\sim$ 4 \footnotemark[6] 
& $\sim$ 1 \footnotemark[7] \\ 
SrMoO$_3$ (${t_{2g}}^2$) & 7.9 \footnotemark[8]
 & $\sim$ 2 & $\sim$ 1 \\ 
\hline
\hline
\end{tabular}
\label{tab1}
\end{center}
${}^a$Ref.~\onlinecite{IHInoue}\hspace{5mm}
${}^b$Ref.~\onlinecite{Sekiyama}\hspace{5mm}
${}^c$Ref.~\onlinecite{Jokamoto}\hspace{5mm}
${}^d$Ref.~\onlinecite{TakizawaSCRO1}

  ${}^e$Ref.~\onlinecite{maenonat}\hspace{5mm}
${}^f$Ref.~\onlinecite{TOguchi}\hspace{5mm}
${}^g$Ref.~\onlinecite{SekiyamaSr2RuO4}\hspace{5mm}
${}^h$Ref.~\onlinecite{nagai}  
\end{table}

The hallmarks of strong correlations such as the band narrowing and
lower Hubbard bands observed in SrVO$_3$ which might be expected to be
even more easily resolvable in the photoemission on ruthenates due to
the larger mass enhancement, thus turn experimentally more elusive to
find. On the other hand, the ruthenates have been also pictured as
influenced substantially by the proximity to the magnetic transition
and the corresponding magnetic fluctuations \cite{JMcao} which brings
to the problem ingredients whose photoemission signatures are
understood less well. 

In this context, it is important and interesting
to study the effects of electron correlation in another
perovskite-type $4d$ oxide SrMoO$_3$, which is a Pauli paramagnetic
metal \cite{paulipara} and is therefore expected to show electron
correlation free from possible effects of magnetism and
 superconductivity.  Molybdates are particle-hole analogues of
ruthenates as far as the occupancy of the $t_{2g}$ shell is concerned:
in SrMoO$_3$ ($4d^2$) the $t_{2g}$ band is occupied by two electrons,
and in SrRuO$_3$ (low-spin $4d^4$) the $t_{2g}$ band is occupied by
two holes.

In this paper, we investigate the correlated density-of-states (DOS)
of SrMoO$_3$ experimentally with photoemission and compare the results
to the band-structure calculations within local-density approximation
(LDA) and the LDA combined with the dynamical mean-field theory
(LDA+DMFT). We also compare the results to the analogous ones on SrVO$_3$. The comparison between
SrMoO$_3$ and SrVO$_3$ can be expected to be particularly revealing.
First, both materials are paramagnets.  Second, both have intermediate
degrees of correlations, judging from the effective mass
($m^{\ast}/m_b$).  Finally these two materials have overall similar
band-structure at the level of local density approximation (LDA) as
shown in Fig.~\ref{fig3}, and a low filling of the $d$-shell: $d^2$
and $d^1$, respectively, which means that unlike ruthenates they are
not affected that much by the possible effects of proximity to the
 van-Hove singularity, see, e.g.~Ref.~\cite{JMmravlje11}.

 Nagai {\it et al.} \cite{nagai} reported that SrMoO$_3$ single
 crystals grown in ultra-low oxygen pressure show resistivity as low
 as 5.1 $\mu\Omega$cm at 300 K.  Recently, Radetinac {\it et al.}
 \cite{SMOfilm} reported the fabrication of high-quality SrMoO$_3$
 thin films using argon gas in the pulsed laser deposition (PLD)
 process. In this study, we fabricated the same type of thin film
 which has atomic-level flatness at the surface (with a
 root-mean-square roughness of 0.2 nm \cite{SMOfilm}) and performed
 hard x-ray (HX) photoemission spectroscopy measurements.  By applying
 bulk-sensitive HX photoemission spectroscopy to the atomically flat
 surface, we succeeded in obtaining the spectrum of bulk Mo $4d$
 bands.  The obtained bulk Mo $4d$ bands do not show band narrowing
 from the band-structure calculations.  This behavior, which is also
 observed for SrRuO$_3$, can be understood if the genuine coherent
 quasiparticle peak exists only near the Fermi level.  This result is
 an experimental manifestation of the recent DMFT with Hund's rule coupling in Ref.~\cite{Anto,Anto1},
 demonstrating that in the case of ${t_{2g}}^2$ or ${t_{2g}}^4$
 systems the overall $d$ bandwidth is not strongly affected by
 electron correlations. 
In the theoretical part of the paper we 
report further DMFT model 
calculations that show explicitly 
that the Hubbard bands are pulled 
to lower energies by the Hund's rule coupling and 
that the spectral weight is redistributed 
within the quasiparticle band instead of 
being shifted to the Hubbard satellites as is the case in SrVO$_3$. 
Similar results have been discussed also for 
iron-based superconductors \cite{kutepov}. 

We also performed realistic LDA+DMFT calculations of SrMoO$_3$. The
calculated $t_{2g}$ spectral functions do not show obvious band
narrowing on large ($\sim$ 1 eV) energy scales, 
yet the low-energy spectra is renormalized, 
in agreement with measurements. Comparison of the theory to 
the measurements reveals additional spectral weight at $\sim -2$ eV 
that is not present in the theory. We argue that this is not the lower 
Hubbard band but rather a plasmon satellite, an interpretation that is 
supported by the existence of satellites in the measured core-level 
spectra. 

The paper is structured as follows. In Sec.~\ref{secExp} we describe
the experimental procedure and in Sec.~\ref{secTheory} we describe
some details of the theoretical work. In Sec.~\ref{secPE} we report
the photoemission results and compare them to what is found in the
band-structure LDA and LDA+DMFT calculations. 
Section \ref{sec:Disc} 
contains discussion of our results and 
in Sec.~\ref{secConc} we conclude. 
In Appendix~\ref{sec_lda_dmft} we give LDA+DMFT 
self-energies and discuss LDA+DMFT results in more
details. Appendix~\ref{secDMFT} contains DMFT results for a model
system that shows explicitly how Hubbard bands are pulled in by the
Hund's rule coupling, which supports our interpretation of the
experimental spectra.

\section{Experiment} \label{secExp}
The SrMoO$_3$ thin film was grown in the (001) direction on a
GdScO$_3$ (110) substrate [($a=5.482\ \mbox{\AA}$,
  $b=5.742\ \mbox{\AA}$, and $c=7.926\ \mbox{\AA}$ for the
  orthorhombic lattice ($\bar{a}=3.967\ \mbox{\AA}$ for a pseudocubic
  lattice definition)] by the PLD method.  Since the lattice constant
of the cubic SrMoO$_3$ is 3.976 $\mbox{\AA}$ \cite{str}, a lattice
mismatch between substrate and film is only $-0.2$ \%.  The thickness
of the thin film was about 70 nm.  The details of the fabrication were
described in Ref.~\cite{SMOfilm}.  

HX photoemission measurements were 
carried out at BL-47XU of SPring-8. 
No surface cleaning was performed 
before the measurements. The HX photoemission spectra were
recorded using a Scienta R-4000 electron energy analyzer with a total
energy resolution of 300 meV at the photon energy of 7.94 keV. 
 We also performed SX photoemission measurements at Photon Factory 
BL-2C to obtain the information about surface states. 
The SX photoemission spectra were 
recorded using a Scienta SES-2002 electron energy analyzer 
with a total energy resolution of 300 meV 
at the photon energy of 780 eV. 
The position of $E_F$ was determined by measuring the spectra of 
gold which was in electrical contact with the sample. 
All the spectra were measured at room temperature.

\section{Theory} \label{secTheory}
 The band structure was calculated using the linearized augmented
 plane wave method implemented in the WIEN2K package. Bulk SrMoO$_3$
 has an orthorhombic crystal structure below 150 K, a tetragonal one
 between 150 K and 250 K, and cubic structure at temperatures above
 250 K~\cite{str}.  The degree of distortions 
 away from cubic symmetry 
 in this compound is small, and we found that the band structure of the compound is not
 influenced by distortions significantly. The results which we report
 below are for the cubic perovskite structure with the lattice
 constant $a=3.976\ \mbox{\AA}$ \cite{str}.  The LDA+DMFT calculations
 were done in the framework described in
 Ref.~\cite{JMaichhorn_prb_2009,JM_TRIQS}. Full rotationally invariant
 interaction with Kanamori parameters $U=3.0$ eV and $J=0.3$ eV has been
 used~\cite{JMvaugier}. {Using these parameters, 
the calculated mass enhancement is $\sim 2$, 
consistent with experiment. 
The analytical continuations of the data to real frequencies 
were performed using stochastic maximum entropy method \cite{beach}. 

To support our interpretation of the results, 
we also carried out the model calculations within DMFT, which were 
done using the semicircular DOS.

\section{Photoemission results; comparison to LDA and LDA+DMFT} \label{secPE}
Figure~\ref{fig1} shows the core-level photoemission spectra of the
SrMoO$_3$ thin film.  The O $1s$ spectrum (a) shows that the
``contamination'' signal on the higher binding-energy side is weak.
The Sr $3d$ spectrum (b) has only one component at $3d_{5/2}$ and
$3d_{3/2}$.  These two results demonstrate that our photoemission
spectra are free from surface degradation or contamination and
represent the bulk electronic properties of the SrMoO$_3$ thin film.
The Mo $3d$ spectrum (c) has two structures at $-229.3$ eV and $-232.5$ eV,
almost the same as those of MoO$_2$ (Mo$^{4+}$)~\cite{MoJAP},
 representing the bulk Mo$^{4+}$ states. 
This core level was also measured at 780 eV in the SX region, 
and two structures were observed at $-233.3$ eV and $-236.4$ eV, 
almost the same as those of MoO$_3$ (Mo$^{6+}$)~\cite{MoJAP}. 
It also has some weak Mo$^{4+}$ signal at $-229.3$ eV. 
These results mean that our thin film had Mo$^{4+}$ in bulk and 
the surface states were dominated by Mo$^{6+}$. 
Since Mo$^{6+}$ has no $4d$ electrons, 
such surface oxidized states do not affect the Mo $4d$ band 
which will appear in subsequent figures. 

There is also some additional intensity 
in all the core levels, that is, 
at $-532.5$ eV in O $1s$ (a),
at $-138$ eV in Sr $3d$ (b), 
and at $-235$ eV in Mo $3d$ (c). 
We plotted these three core levels 
as a function of relative energy 
to the main peak 
in Fig.~\ref{fig1} (d). 
One can see the intensity 
around $-2$ eV in all the core levels, 
which points to a common 
origin. As we discuss in more detail later, we believe 
that these structures 
are due to plasmon satellite. The 
energy of $\sim 2$ eV coincides closely with the plasma 
frequency reported 
in the measurements of reflectivity \cite{pla}. 

\begin{figure}
\begin{center}
\includegraphics[width=8cm]{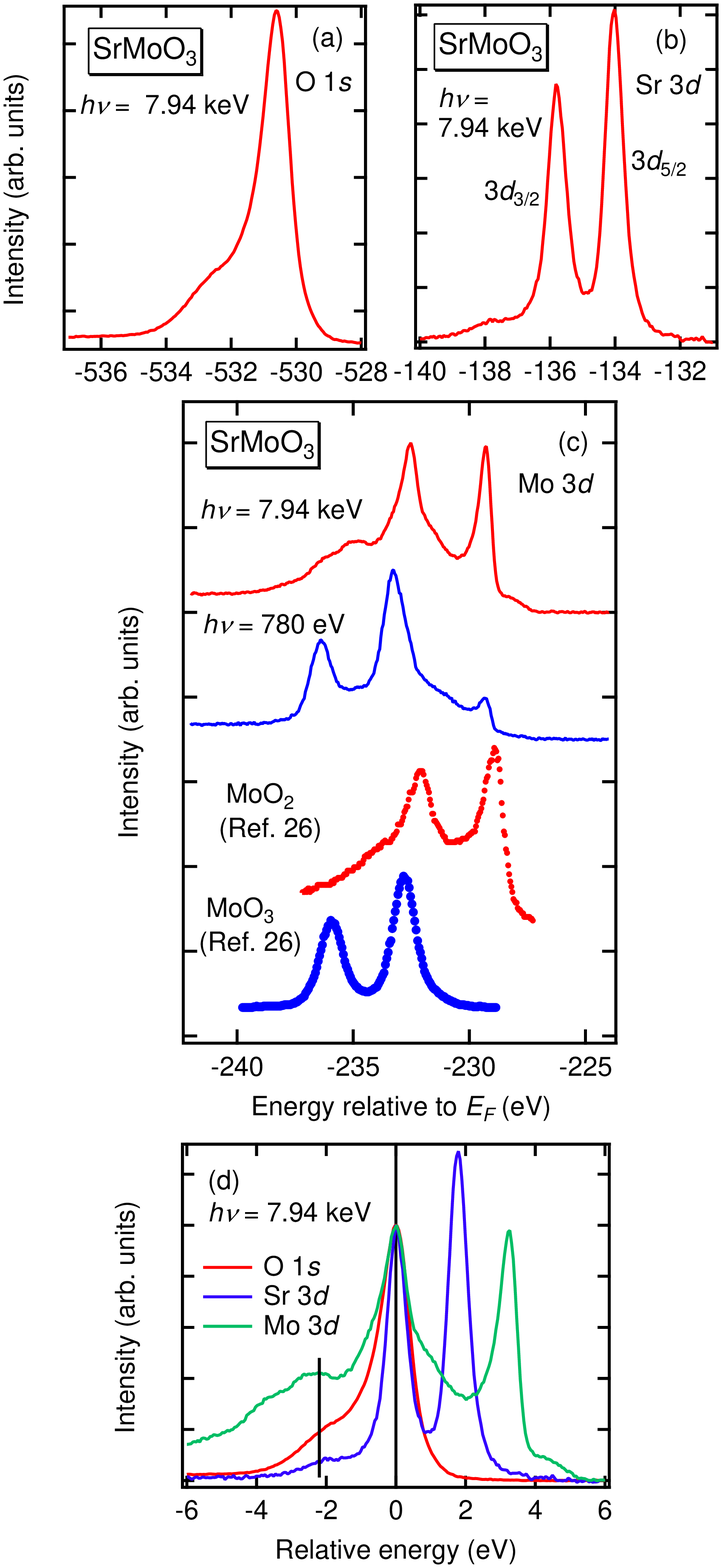}
\caption{(Color online): 
Core-level photoemission spectra of 
the SrMoO$_3$ thin film. 
(a) O $1s$. (b) Sr $3d$. (c) Mo $3d$. 
Mo $3d$ spectra were measured by both HX and SX, 
plotted together with the reference spectra 
of MoO$_2$ and MoO$_3$ (Ref.~\onlinecite{MoJAP}). 
Panel (d) shows all the core levels plotted 
as a function of relative energy to the main 
peak.}
\label{fig1}
\end{center}
\end{figure}

Figure~\ref{fig2} (a) shows the valence-band photoemission spectrum of
the SrMoO$_3$ thin film.  By comparing the photoemission spectrum with
the DOS deduced from the LDA (b), one can see that the Mo $4d$ band
is located near $E_F$, and the O $2p$ band is located on the
higher-energy side (from $-4$ to $-10$ eV).  The dashed line in panel
(a) show the tail of the O $2p$ band extended towards the Mo $4d$
band.  One can also clearly see the Mo $4d$ band crossing $E_F$ and 
that the photoemission signal is described well 
by the LDA DOS.  

\begin{figure}
\begin{center}
\includegraphics[width=9cm]{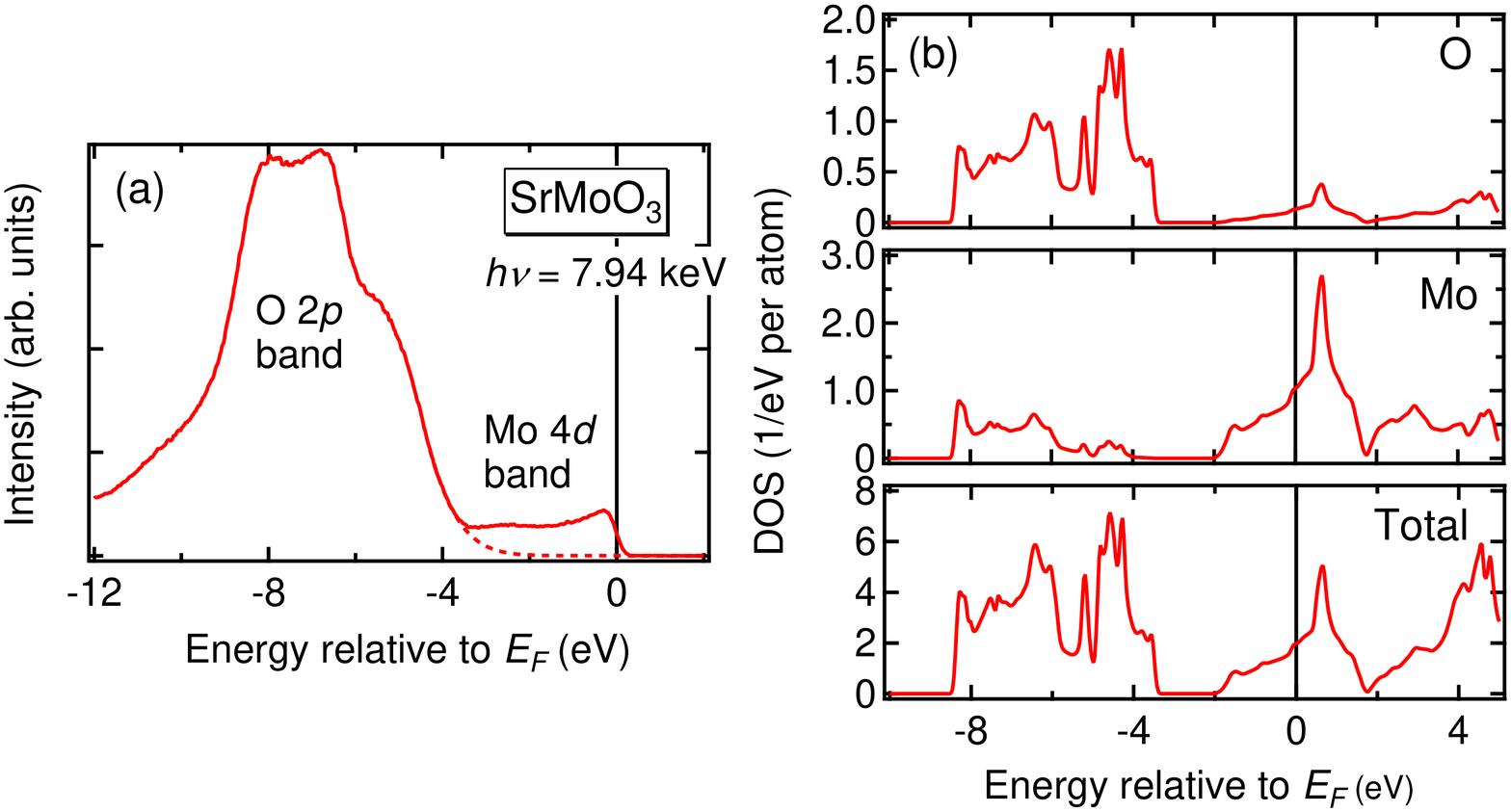}
\caption{(Color online): Electronic structure of SrMoO$_3$ 
near the Fermi level. 
(a) Valence-band photoemission spectrum of a 
SrMoO$_3$ thin film. 
The dashed line shows the estimated tail of the O $2p$ 
band. (b) The DOS of SrMoO$_3$ obtained 
from band-structure calculations.}
\label{fig2}
\end{center}
\end{figure}

Figure~\ref{fig3} (a) shows the bulk $d$ orbital component obtained by
subtracting the oxygen contribution (dashed line in
Fig.~\ref{fig2} (a)) from the photoemission signal. This photoemission
  DOS is compared to the results from the LDA band-structure
  calculation for SrMoO$_3$ as well as to the $t_{2g}$ DOS from the
  LDA+DMFT calculation, which includes the effects of correlation.
  The results for SrVO$_3$ are also shown in panel (b), where the
  experimental spectrum is taken from Ref.~\cite{Sekiyama}. The
  calculated DOS has been broadened with a Gaussian of 0.3 eV (FWHM: a
  full width at half maximum) and an energy-dependent Lorentzian
  ($\mbox{FWHM}=0.2|E-E_F|$ eV) \cite{Hufner} to account for the
  instrumental resolution and the lifetime broadening of the
  photohole, respectively. The theoretical data was multiplied by the
  Fermi function. 

The first observation is that the experimental photoemission is
distinct from the LDA results, which is a signature of electronic
correlations. In the case of SrVO$_3$, these manifest in an obvious,
well-known way: the quasiparticle band is narrowed and a split-off
lower Hubbard band is seen. These results are reproduced very well by
the LDA+DMFT.

In the case of SrMoO$_3$, the electronic correlations manifest in a 
different way. The observed photoemission does not show band 
narrowing. Rather, the quasiparticle band appears 
to be widened and develops a hump at energy $-2.5$ eV. 

The LDA+DMFT data also does not display an obvious band narrowing and
is consistent with experiment  above -1eV.  Whereas part of
the spectral weight extends also to frequencies below $-2$ eV due to
the large imaginary part of the self energy found in this energy range
(for the self energies as well as about the spectra at positive
frequencies, see Appendix~\ref{sec_lda_dmft}), our LDA+DMFT
calculations cannot account for the experimental excess weight at
$\sim -2.5$ eV. This leads us to propose that the hump is not a
Hubbard band. Note also that the hump occurs at an energy which is
separated more from the Fermi energy than the Hubbard band in
SrVO$_3$, whereas the interaction parameters are expected to be
smaller for $4d$ elements than for $3d$ elements due to the more
extended orbitals of the former. Weaker electronic correlation in $4d$ TMOs 
CaRuO$_3$ and SrRuO$_3$ than $3d$ TMOs 
were reported in Ref.~\onlinecite{kmaiti} 
from the comparison between photoemission spectra and 
first-principles calculations. 

What is then the origin of the hump? Whereas more work will be needed
to clarify this conclusively, we believe the most natural explanation 
is that it is a plasmon satellite. The plasma edge in the optical 
experiments~\cite{pla} is indeed at about 2 eV, and we stress again 
that the satellite structures are seen also in all the core levels. 
Influence of plasmons in core-level photoemission spectra 
was observed in simple metals like Mg \cite{ley,leybook} and 
conducting oxides like Na$_x$WO$_3$ \cite{wo3} and K$_{0.3}$MoO$_3$ \cite{moo3}. 
In correlated materials the influence of plasmons on
photoemission is discussed less often. We note that the plasma
frequencies are about 2 eV, which means that they often overlap with
the Hubbard band and that therefore $4d$ oxides that do not show
pronounced Hubbard bands might be promising materials
to investigate plasmons further. Molybdates, that have a gap in the
LDA spectrum between $-3$ and $-2$ eV are particularly promising in this
respect. 

\begin{figure}
\begin{center}
\includegraphics[width=9cm]{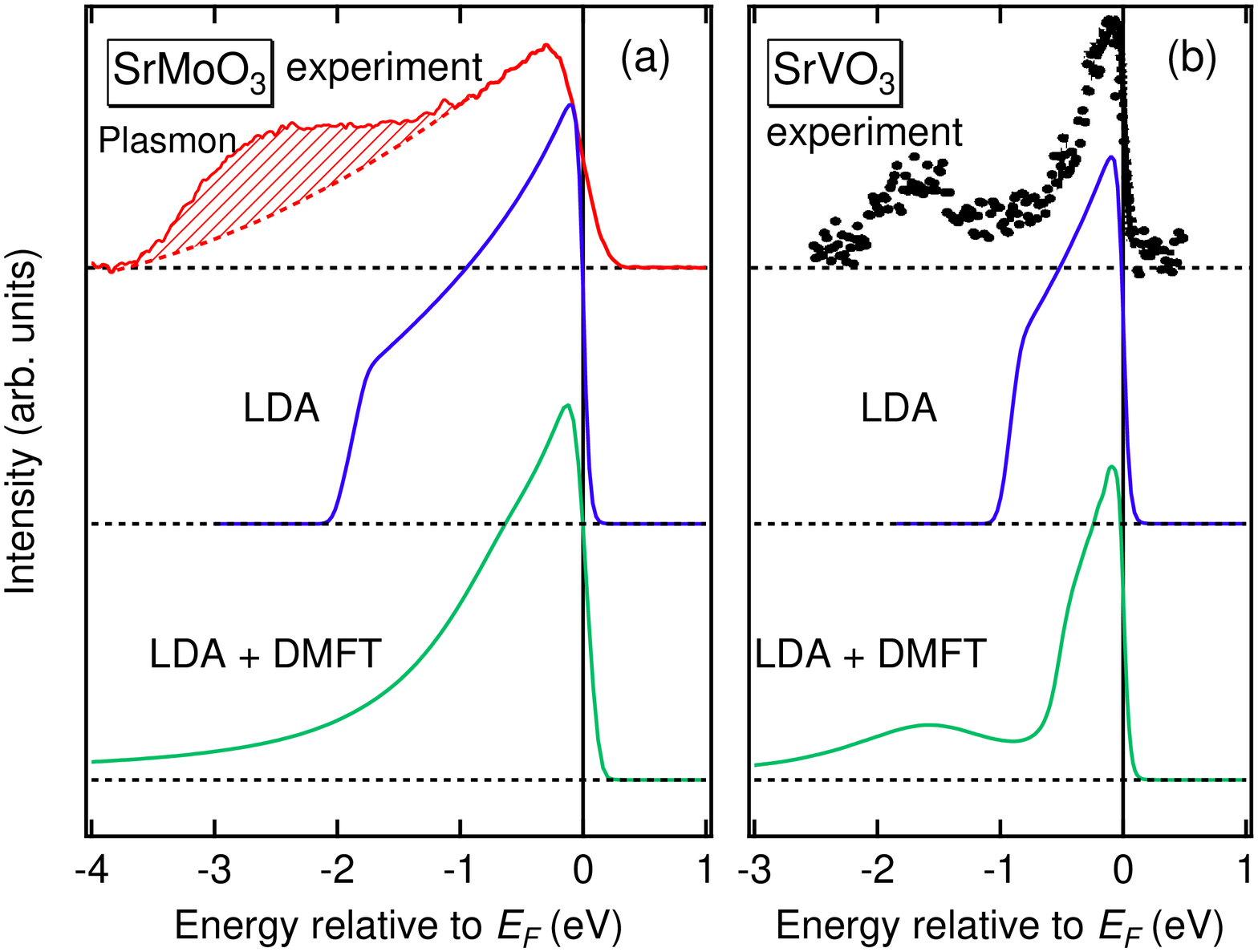}
\caption{(Color online): Comparison of the bulk component obtained by 
photoemission spectroscopy and the band-structure calculation 
for SrMoO$_3$ (a) and SrVO$_3$ (b). The experimental spectrum 
of SrVO$_3$ is from Ref.~\cite{Sekiyama}. 
The calculated DOS 
has been broadened with a Gaussian and a Lorentzian function. 
The hump structure at $-2.5$ eV in experiment (a) is 
attributed to a plasmon satellite.}
\label{fig3}
\end{center}
\end{figure}

\section{Discussion} 
\label{sec:Disc}
We now turn to the origin of the difference observed in photoemission between 
SrMoO$_3$ and SrVO$_3$.  Why does the latter exhibit a narrowed
quasiparticle band and the Hubbard satellites and the latter not?

Recently, de' Medici {\it et al.}  \cite{Anto,Anto1} calculated the 
quasiparticle weight $Z$ in $t_{2g}$ systems by including the effects 
of Hund's rule coupling $J$.  The results for ${t_{2g}}^1$ and
${t_{2g}}^2$ are reproduced in Fig.~\ref{fig4}, where $U$ is the on-site
Coulomb interaction and $D$ is the half bandwidth.  There is a marked
difference between the ${t_{2g}}^1$ and ${t_{2g}}^2$ systems.  In the
case of the ${t_{2g}}^1$ system, a non-zero $J$ increases $Z$, whereas
in the case of the ${t_{2g}}^2$, ${t_{2g}}^4$ systems a non-zero $J$
rather decreases $Z$ (unless $U/W$ is too large). 
This means that in the latter case even small 
values of $U/W\lesssim 1$ may lead to a suppressed $Z$ at small
energy scales. Because so small values of $U$ are sufficient 
to suppress $Z$, the
high energy scales cannot be separated from the small energy scales
in a clear way and instead of the narrowing of the quasiparticle
band, the spectral weight is redistributed within it. 
This is illustrated in more details by DMFT calculations on a model system in 
Appendix~\ref{secDMFT}. 

\begin{figure}
\begin{center}
\includegraphics[width=9cm]{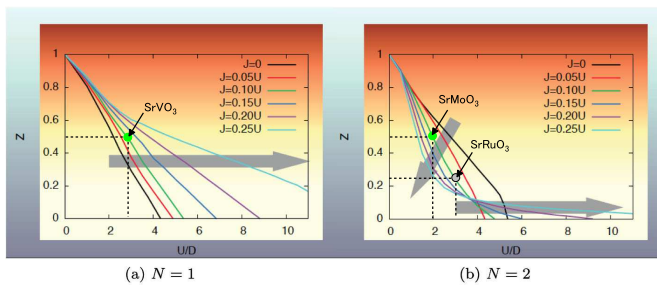}
\caption{(Color online): Quasiparticle weight $Z$ as a function of 
$U/D$ for $N=1$, 2 electrons in three orbitals \cite{Anto}. 
The gray arrows indicate the influence of 
an increasing Hund's rule coupling $J/U$. 
SrRuO$_3$, plotted in $N=2$, is actually a $d^4$ occupancy 
and one should note that 
there is not an electron-hole symmetry for a $t_{2g}$ DOS.}
\label{fig4}
\end{center}
\end{figure}

In the case of SrMoO$_3$, we obtain $\gamma_b=4.6$ mJ/mol K$^2$ 
from the band-structure calculation, 
and $\gamma=7.9$ mJ/mol K$^2$ from specific heat measurements \cite{nagai}. 
These values give $m^{\ast}/m_b=\gamma/\gamma_b\sim 2$, but 
this effect of electron correlation is not clearly observed in 
the photoemission spectra shown in Fig.~\ref{fig3} (a), 
indicating that the genuine coherent 
part with band narrowing exists only near $E_F$ as also 
suggested for SrRuO$_3$ \cite{TakizawaSCRO1}. 
In the case of ${t_{2g}}^2$ (or ${t_{2g}}^4$) systems, 
nonzero $J$ can decrease $Z$, which is opposite to 
the case of ${t_{2g}}^1$. Small values of $Z$ 
is not necessarily realized  
by a large value of $U/D$. 
Therefore, the overall $d$ bandwidth, determined by the value of $U/D$, 
can be rather free from electron correlation effects. 

Since $Z=(\gamma/\gamma_b)^{-1}$, if we assume $J=0.10U$, we obtain
$U/D\sim 3$ for SrVO$_3$, and $U/D\sim 2$ for SrMoO$_3$.  One can
clearly see that SrVO$_3$ has a larger value of $U/D$ and is closer to
the Mott transition than SrMoO$_3$.  
This makes the case for SrMoO$_3$ being a ``Hund's metal'' with intermediate correlations 
and explains the difference in spectra between the two materials shown in Fig.~\ref{fig3}.

\section{conclusion} \label{secConc}
We have studied the electronic structures of a SrMoO$_3$ thin film by
HX photoemission spectroscopy. From the Mo $3d$ core level, we found
that the valence of Mo is indeed $4+$ in the bulk.  The valence-band
spectrum clearly showed the Mo $4d$ band crossing $E_F$.  We compared
thus obtained bulk Mo $4d$ band with the band-structure
calculation.  Although the electronic specific heat shows a clear
signature of mass enhancement, we did not observe a band narrowing
effect arising from electron correlation effects, suggesting that the
genuine coherent part exists only near $E_F$. 
Such a behavior was also
observed in SrRuO$_3$, and is considered to be an experimental
manifestation of the recent DMFT studies including Hund's rule
coupling \cite{Anto,Anto1}. 
The LDA+DMFT cannot account for the broad 
hump at $\sim -2.5$ eV. The relatively high-binding energy of the 
feature as well as discrepancy with the LDA+DMFT suggests the hump is 
not a lower Hubbard band. We believe the hump is a plasmon satellite 
and propose that molybdates might be an interesting material to 
investigate physics of plasmons in correlated materials further. 

\begin{acknowledgements}
This research was granted by the Japan Society for the Promotion of 
Science (JSPS) through the 
``Funding Program for World-Leading Innovative R\&D on Science 
and Technology (FIRST Program),'' 
initiated by the Council for Science and Technology Policy (CSTP). 
 This work was supported in part by JSPS Grant-in-Aid 
for Scientific Research(S) No. 24224009. 
The synchrotron radiation experiments at SPring-8 were 
performed under the approval of the Japan Synchrotron 
Radiation Research Institute 
(2010B1740 and 2011A1624). 
The synchrotron radiation experiments at KEK-PF were 
performed under the approval of the Program 
Advisory Committee (Proposal 
No. 2008S2-003) at the Institute of Materials 
Structure Science, KEK. The computing time was provided by IDRIS-GENCI
under grant 2011091393. JM acknowledges support 
of the Slovenian Research Agency (ARRS) under Program 
P1-0044.
\end{acknowledgements}
\appendix
\section{LDA+DMFT DOS and self energies} \label{sec_lda_dmft}
 For SrVO$_3$ the values of interaction $U=4.5$ eV and
 $J/U=0.15$ was used in the calculation. Note that these are the
 screened values, compatible with the $t_{2g}$ only calculation that
 we do here. Starting from the same ratio of atomic interaction
 parameters, $J_{\mathrm{atom}}/U_{\mathrm{atom}}$ the screened values
 of $J/U$ can be expected to be higher in $3d$ oxides, because the
 screening which is stronger in $3d$ oxides due to the proximity of the
 oxygen states affects more $U$ than the $J$ which is related to
 higher orders of multipole expansion of the Coulomb interaction.

 On top panels of Fig.~\ref{fig:sigma} the LDA+DMFT DOS are compared
 to the LDA DOS for SrMoO$_3$ (data shown left) and SrVO$_3$
 (right). In the case of SrVO$_3$, $U/W\sim 2>1$, hence the split-off
 Hubbard bands are found. In the case of SrMoO$_3$, $U/W\lesssim 1$,
 hence the high energy features are merged with the quasiparticle
 band. More pronounced difference as well as a precursor of the Hubbard
 band at about $1.5$ eV are actually found on the positive frequency
 side which is invisible to photoemission experiments.

 On medium panels of Fig.~\ref{fig:sigma} the real and on bottom
 panels the imaginary parts of the self energy $\Sigma$ are shown. For
 SrMoO$_3$, on the negative side, the self energy deviates from the
 linear behavior at about $-0.5$ eV and levels off to the frequency
 independent behavior, recovering the bare LDA dispersion, although
 with significant broadening (Im$\Sigma$, shown in lower panel) below that
 frequency. Conversely, for SrVO$_3$ the imaginary part of the self
 energy shows a weak pole-like structure, signifying the Hubbard band,
 and related feature with positive slope in the real part of the self
 energy in the same energy range.

 The magnitude of Im$\Sigma$ is found to be comparable at small
 frequencies (note the data is described well by a parabola with
 similar curvature). At higher frequencies $\sim 1$eV, on the other hand, the
 magnitude of Im$\Sigma$ in SrMoO$_3$ becomes relatively smaller, which is
 another manifestation of the fact that the electronic correlations
 originate from the multiplet splittings which become unimportant at
 frequencies above $J$.

\begin{figure}
 \begin{center}
   \includegraphics[width=1.0\columnwidth,keepaspectratio]{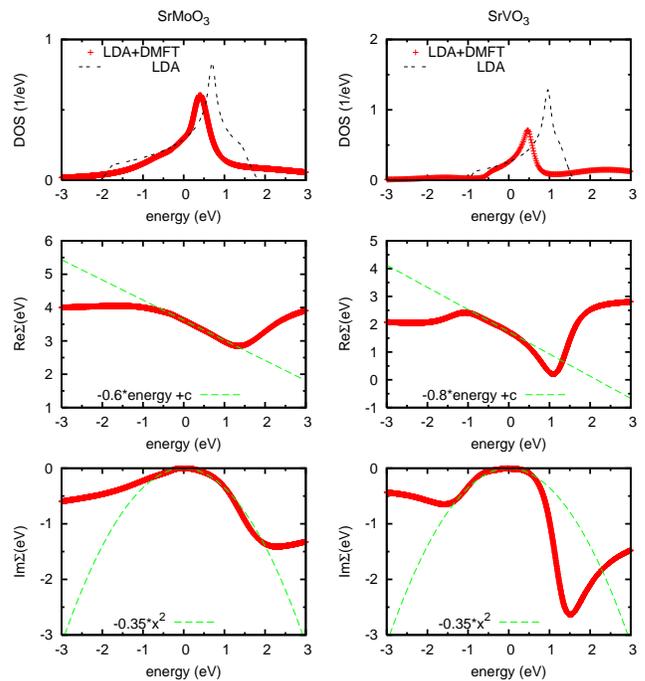}
   \end{center}
   \caption{\label{fig:sigma} Top panels: LDA and LDA+DMFT DOS for
     SrMoO$_3$ (left) and SrVO$_3$ (right). Medium panels: Real parts
     of self energy. Linear fit is also shown. Bottom panels:
     Imaginary parts of the self energy. Parabolic fit is also shown.}
 \end{figure}

\section{Model DMFT results}
\label{secDMFT}
To further support the interpretation of the absence of Hubbard bands 
in terms of the effects of the Hund's rule coupling,  we calculated the
correlated DOS for a model with semicircular DOS for several ratios of
$J/U=0.0$, 0.05, 0.1, 0.15 
(with $J/U=0.15$ being close to the physical
values for the transition-metal oxides), adjusting $U$ so that the low
frequency mass renormalization remains constant 
$m^{\ast}/m=Z^{-1}\approx 4$.

\begin{figure}
\begin{center}
\includegraphics[width=1.0\columnwidth,keepaspectratio]{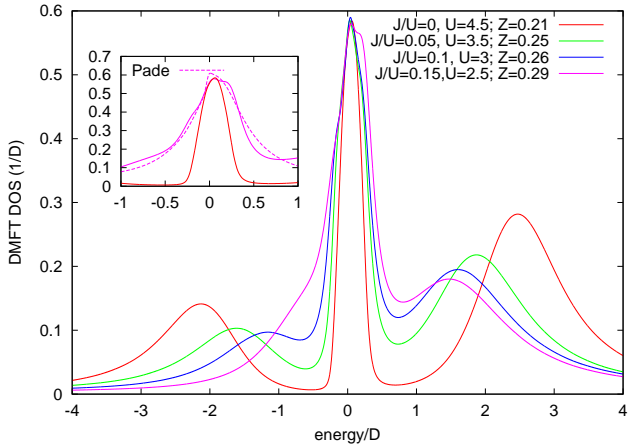}
\end{center}
\caption{\label{fig:t2g_spectra}The DMFT DOS for the $t_{2g}$ model
     with semicircular noninteracting DOS occupied by 2
     electrons/atom. Several values of $J/U$ were used and $U$ is
     adjusted so that the quasiparticle renormalization is close to
     $Z=1/4$. Inset: close-up on quasiparticle peak for $J/U=0,0.15$. For $J/U=0.15$, also the Pad\'e data is 
      shown for comparison. Other curves are obtained using maximum entropy analytical continuation. }
\end{figure}

The calculated DOS are shown on Fig.~\ref{fig:t2g_spectra}. 
At $J/U=0$, for sizable correlations to occur, the interaction
strength must me tuned quite close to the Mott transition (which
occurs for $U/D\approx 5.5$). The resulting spectra are akin to the
spectra of doped Mott insulators: pronounced Hubbard bands and the
narrow quasiparticle peak, implying clear separation of high and low
energy scales.

As $J/U$ is increased, two effects occur. First, as the correlations
due to $J$ develop a smaller value of $U$ is sufficient to reach the
same degree of renormalization. Second, the effective atomic
interaction for a transfer of electron, given by $E(N+1)+E(N-1)-2 E(N)$ that is equal to $U-3J$ for a $t_{2g}$ atom away from half-filling, 
is diminished even more with increasing $J/U$, thus the Hubbard bands
are pulled in by the Hund's rule coupling $J$.

This is reflected in the spectra by the diminishing of the
peak-to-peak distance between the Hubbard bands 
(4.7, 3.7, 2.7, 2.1 in
units of $D$ for increasing $J/U=0.0, ...,0.15$, respectively). At the
largest $J/U=0.15$ considered, the lower Hubbard band starts to
overlap with the quasiparticle band and remains visible only as a mild
shoulder. The upper Hubbard band, however, remains visible.  The
high-energy and the low-energy scales are not separated clearly
anymore, the incoherent spectral weight is transfered to energies
which overlap with the quasiparticle band and therefore influence its
shape. The quasiparticle peak appears broader and obtains an
asymmetric shape. In the inset, we replot the DOS in the narrower
frequency range. The quasiparticle peak at $J/U=0.15$ has markedly
different shape from that of the $J/U=0$ case, whose shape closely resembles
that of the narrowed noninteracting (semicircular) DOS.
Next to the stochastic maximum entropy analytical continuation we show
also the data obtained by the Pad\'e approximants. The broadened
quasiparticle peak with asymmetric shape is seen from both analytical
continuations, which therefore is likely a genuine feature of Hund's
metals and which deserves further exploration.

On Fig.~\ref{fig:t2g_self} we show the self energies for $J/U=0$ and
$J/U=0.15$.  The real part of the self energies exhibit similar low
frequency slope, which corresponds to the matching quasiparticle
residue $Z$, but in other aspects the data for $J/U=0.15$ differs
substantially from the data at vanishing Hund's coupling strength. The
$J/U=0$ real part of the self energy follows a quasi-linear dependence
up to a high energy scale followed by an abrupt feature indicating the
onset of the Hubbard band. Conversely, for $J/U=0.15$ the real part of
the self energy is linear only up to a small frequency scale.  At
higher frequencies relatively mild frequency dependence is seen, which
indicates a weaker overall band narrowing. Except on approaching the
Hubbard bands, the magnitude of Im$\Sigma$ (bottom inset) for
$J/U=0.15$ is larger, indicating correlations that develop due to the
Hund's rule coupling despite a significantly smaller value of $U$.
The genuine coherent, but strongly renormalized part, is thus actually
limited to the low frequency scales ($\lesssim 0.1D$ for the present
data), whereas at higher frequency scales a larger dispersion, but
with much shorter lifetime is recovered.

Interesting further insight into the particular behavior of self
energies for Hund's metals is obtained by considering Kramers-Kronig
relations. Take that Im$\Sigma = -A \omega^2$ up to a cutoff
$\omega_c$, and that it vanishes at $|\omega| > \omega_c$.
Kramers-Kronig relations give Re$\Sigma = -2 \omega_c A \omega +\cdots$. The
slope of Re$\Sigma$ which determines the quasiparticle renormalization
increases both with $\omega_c$ and $A$. At a fixed quasiparticle
renormalization, this relation also tells that the cutoff frequency
(which has the meaning of the energy of the kink) and the curvature
are related.  For vanishing Hund's rule
coupling, the curvature of Im$\Sigma$ is small but persists up to a
larger cutoff. For physical values of $J$, the curvature is larger, but it holds
only up to a small frequency scale. This explains why in Hund's
metals, like ruthenates, the kinks are often found (see,
e.g.~\cite{JMmravlje11} and references therein) at small energy
scales. 

\begin{figure}
 \begin{center}
   \includegraphics[width=1.0\columnwidth,keepaspectratio]{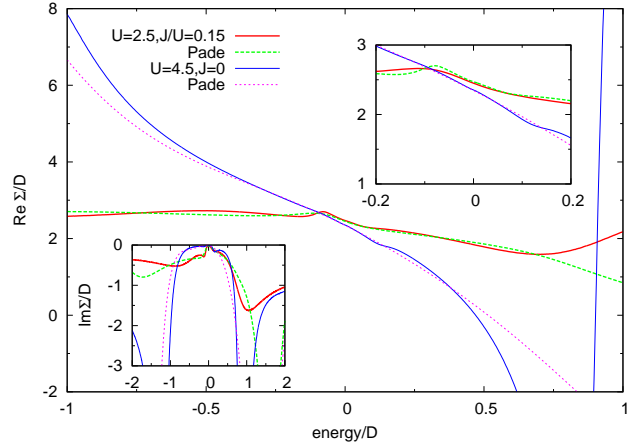}
   \end{center}
   \caption{\label{fig:t2g_self}The DMFT self energies for the
     $t_{2g}$ model for $J/U=0$ and $J/U=0.15$. Analytical
     continuations using Pad\'e approximants and maximum entropy
     method are shown. The
      $J/U=0$ results are shifted vertically for easier comparison between the two datasets.
  Top inset: close-up to low frequencies.  Bottom inset: imaginary part of self energies.
 }
 \end{figure}

 \bibliography{LVO1tex}

\end{document}